# Elastic Properties of Amorphous $LiTaCl_6$ Solid-State Electrolyte

Xiaolin Liu[1] and De-en Jiang[1, *]

[1] Department of Chemical and Biomolecular Engineering, Vanderbilt University, Nashville, Tennessee 37235, USA

*E-mail: de-en.jiang@vanderbilt.edu

**Abstract:**

Amorphous solid-state electrolytes are attractive candidates for safe, high-energy-density all-solid-state batteries, yet their mechanical properties remain poorly understood from a computational perspective. Here, we investigate the elastic behavior of the recently discovered amorphous superionic Li-ion conductor $LiTaCl_6$ using density-functional-theory (DFT)-based methods, including unrelaxed static, relaxed static, and strain fluctuations from molecular dynamics (MD) simulations in the isobaric–isothermal (*NPT*) ensemble with DFT-trained machine-learning force fields (MLFFs). While the unrelaxed static method predicts a Young's modulus an order of magnitude higher than the experiment, the relaxed static method—commonly applied to crystalline electrolytes—still overestimates the modulus by more than 170%. In contrast, the MD approach using MLFFs yields a Young's modulus of $2.84 \pm 0.26$ GPa, which quantitatively agrees with the experimental value of $2.91 \pm 0.32$ GPa. Using the MLFF-MD approach, we further predict bulk modulus (4.44 GPa), shear modulus (1.02 GPa), and Poisson's ratio (0.39) for amorphous $LiTaCl_6$ and conclude that elastically it behaves like a soft polymer or gel. These results demonstrate that amorphous superionic materials possess some unique elastic properties and that, among the methods examined, only the MLFF–MD approach yields quantitative agreement with experiment, highlighting the necessity of a dynamical treatment to simulate their elastic response, consistent with recent findings for crystalline superionic conductors.

## I. INTRODUCTION

Solid-state electrolytes (SSEs) are promising candidates to replace conventional organic liquid electrolytes, offering enhanced safety, wider electrochemical windows, and the potential to enable lithium metal anodes, thereby increasing energy density [1–3]. Superionic materials are solids in which at least one ionic species exhibits liquid-like mobility, typically corresponding to ionic conductivities exceeding 1 mS/cm at room temperature, making them promising SSE candidates. State-of-the-art SSEs can rival commercial liquid electrolytes in ionic conductivity; for example, some sulfide electrolytes exhibit conductivities exceeding 10 mS/cm [4]. Most SSEs reported to date are polycrystalline, where grain boundaries can facilitate dendrite formation [5]. In contrast, grain-boundary-free amorphous SSEs offer an attractive advantage over polycrystalline ones for all-solid-state batteries (ASSBs).

In addition to high ionic conductivity, commercial SSEs should also possess favorable mechanical properties. A high shear modulus is considered necessary to suppress lithium dendrite growth [6], while high ductility is advantageous for manufacturing [7]. Previous studies have investigated the elastic properties of polycrystalline alkali SSEs using density functional theory (DFT) with finite lattice distortions and strain–stress fitting approaches [8,9]. However, conventional static approaches rely on the harmonic approximation, in which ions vibrate around well-defined equilibrium positions, an assumption that generally breaks down in superionic conductors, where a large fraction of ions exhibits liquid-like motion rather than harmonic vibrations. In such cases, static stress–strain approaches implicitly and incorrectly assume that all atoms contribute elastically, leading to a systematic overestimation of stiffness [10]. In addition, non-affine deformation under applied strain, which is ubiquitous in amorphous materials, is not adequately captured by static methods [11]. Computational studies of elastic properties have been rarely performed on amorphous electrolytes, and the applicability of methods developed for crystalline materials is unclear. Given the enhanced structural disorder and non-affine deformation in amorphous systems, the need for dynamical treatments highlighted in Ref. [10] is expected to be even more critical here.

Amorphous halides have gained great interest in the past several years as Li-ion and Na-ion SSEs. For example, experimental studies on amorphous $LiTaCl_6$ reported a high ionic conductivity of 11 mS/cm and a low activation energy of 0.165 eV at room temperature [12,13]. This material has also been used to fabricate all-solid-state Li batteries [12,14]. In addition, amorphous $LiNbCl_6$ [12,15], $NaTaCl_6$ [16], $Li_2ZrCl_6$, and $Li_2HfCl_6$ [17] have also been explored recently. These studies focus on transport properties and do not report full elastic moduli. Therefore, it is highly desirable to be able to accurately predict elastic properties for these amorphous halide electrolytes. In this work, we use amorphous $LiTaCl_6$ as a prototypical case and investigate its elastic properties employing three different computational approaches. While Young's modulus of amorphous $LiTaCl_6$ has been measured experimentally [14], the shear modulus, bulk modulus, and Poisson's ratio have not been reported; their experimental determination relies on acoustic [9] or optical [18] techniques. Atomistic simulations therefore provide a complementary route to obtain intrinsic elastic properties under controlled conditions and to enable systematic comparison across materials. Although we compute only elastic properties, we note that mechanical properties beyond the elastic regime such as ductility [19], fragility [20], shear viscosity [20], and hardness [21–23] can be predicted using elastic properties. We demonstrate that only the molecular dynamics method yields quantitative agreement with the experiment.

## II. COMPUTATIONAL DETAILS

The amorphous structures were generated following the melt–quench procedure of Ref. [13], detailed in Fig. S1 in Ref. [24]. A DFT-optimized $Li_2TaCl_6$ structure (Fm-3m) from the Open Quantum Materials Database (OQMD) [25] was converted to $LiTaCl_6$ by removing four Li atoms. A 2×2×2 supercell (256 atoms) was melted at 1500 K using the NVT ensemble in *ab initio* molecular dynamics (AIMD) for 50 ps. The equilibrium volume was determined by computing the internal energy as a function of a uniform lattice-scaling factor using several NVT simulations, and the volume corresponding to the minimum energy was selected as the reference cell for subsequent calculations. A short *NPT* simulation at 1500 K was performed to validate this equilibrium density. The liquid was then quenched to 300 K using stepwise temperature reductions corresponding to a cooling rate of $1.5 \times 10^{13}$ K/s. The production trajectory at 300 K was carried out in the NVT ensemble, with an additional short *NPT* run used only to confirm that the resulting density was reasonable.

Ten random snapshots were extracted from the last 150-ps 300 K NVT trajectory for use in the static elastic calculations [13]. The elastic properties of the amorphous electrolyte were computed using three DFT-based approaches: finite-strain methods applied to DFT-optimized structures, performed either with fixed internal coordinates (unrelaxed static) or with internal relaxation (relaxed static), and strain-fluctuation analysis from isothermal–isobaric molecular dynamics (*NPT* MD) employing DFT-trained machine-learning force fields. For all three methods, the Vienna Ab-Initio Simulation Package (VASP) [26–28] was used, based on the projector augmented-wave method [29]. We employed the Perdew-Burke-Ernzerhof generalized-gradient functional [30], a plane wave cutoff of 340 eV with Grimme's D3 dispersion correction [31–33], and Γ-only sampling in all calculations, following our previous work on $LiTaCl_6$ [13]. The electronic convergence criterion is set to $10^{-5}$ eV and in relaxed calculations, the forces are converged to below 0.02 eV/Å.

We train a machine-learning force field (MLFF) [34–36] using structures sampled from a 30-ps AIMD simulation starting from one melt-quench structure (Fig. S1). To generate training data, we performed a sequence of three 10-ps constant-pressure AIMD simulations with linearly increasing temperature segments: 300–400 K, 400–500 K, and 500–600 K. This stepwise heating protocol was used to enhance configurational sampling while maintaining the accuracy of *ab initio* force evaluations. The resulting temperature profile is shown in Fig. S2 [24]. Moreover, as measured by the Lechner–Dellago neighbor-averaged bond-orientational order ($\bar{q}_6$) [37] of the Ta sublattice, the 30-ps AIMD simulation training data covers a large range of $\bar{q}_6$ (Fig. S3 [24]) from a liquid-like value of 0.15 to more ordered structures of 0.25 in $\langle \bar{q}_6 \rangle$. We then take 50 snapshots at every 1ps from a 50-ps MLFF simulation at 300 K to validate the accuracy of the MLFF. The training and test errors are at 1 meV/atom and 60 meV/Å for energies and forces, respectively (Fig. S4 and Table S1 [24]).

The production MLFF–MD simulations were performed in the *NPT* ensemble at 300 K and 1 bar using a cubic 2048-atom supercell. Three independent trajectories of 12 ns each were propagated using the same initial structure from the end of the MLFF training but different initial velocities, with the first 8 ns used for equilibration and the final 4 ns used for the elastic-property analysis; their distinctness has been tracked by the time-dependent $\bar{q}_6$ distribution and $\langle \bar{q}_6 \rangle$. The transferability of the trained MLFF to other possible amorphous structures of $LiTaCl_6$ was tested on a more disordered initial structure, generated from purely geometric random packing of the $[TaCl_6]^-$ octahedra with $Li^+$ ions occupying random interstitial sites and as measured by $\langle \bar{q}_6 \rangle$ of the Ta sublattice (see ref. [24]for details).

**A. Static methods**

In both unrelaxed and relaxed static approaches, we apply distortions to the lattice vectors and fit the energy-strain curve to extract elastic tensors [38,39]. The difference between the two is that in the unrelaxed static method, the fractional coordinates are frozen, while in the relaxed static method, they are allowed to relax. Due to the cubic symmetry of the simulation cell, only three kinds of distortions are required to fit the three elastic tensor elements, namely, $C_{11}$, $C_{12}$, and $C_{44}$ [40,41]. In Voigt notation where the full fourth-rank elastic stiffness tensor is represented as a symmetric 6×6 matrix, the stiffness tensor **C** takes the form:

$$\mathbf{C} = \begin{bmatrix} C_{11} & C_{12} & C_{12} & 0 & 0 & 0 \\ C_{12} & C_{11} & C_{12} & 0 & 0 & 0 \\ C_{12} & C_{12} & C_{11} & 0 & 0 & 0 \\ 0 & 0 & 0 & C_{44} & 0 & 0 \\ 0 & 0 & 0 & 0 & C_{44} & 0 \\ 0 & 0 & 0 & 0 & 0 & C_{44} \end{bmatrix}. \quad (1)$$

The distorted lattice vectors $\mathbf{A}'$ are [42]:

$$\mathbf{A}' = \mathbf{A} \cdot (\mathbf{I} + \boldsymbol{\epsilon}), \quad (2)$$

where **A** are the original lattice vectors:

$$\mathbf{A} = \begin{pmatrix} \boldsymbol{a} \\ \boldsymbol{b} \\ \boldsymbol{c} \end{pmatrix}, \quad (3)$$

**I** is identity matrix and $\boldsymbol{\epsilon}$ is the applied distortion

$$\boldsymbol{\epsilon} = \begin{bmatrix} \epsilon_1 & \frac{\epsilon_6}{2} & \frac{\epsilon_5}{2} \\ \frac{\epsilon_6}{2} & \epsilon_2 & \frac{\epsilon_4}{2} \\ \frac{\epsilon_5}{2} & \frac{\epsilon_4}{2} & \epsilon_3 \end{bmatrix}. \quad (4)$$

The energy-strain relationship under the harmonic approximation is

$$\Delta E = \frac{V_0}{2} \tilde{\boldsymbol{\epsilon}}^{\mathrm{T}} \mathbf{C} \tilde{\boldsymbol{\epsilon}}, \quad (5)$$

where $\tilde{\boldsymbol{\epsilon}}^T = (\epsilon_1 \quad \epsilon_2 \quad \epsilon_3 \quad \epsilon_4 \quad \epsilon_5 \quad \epsilon_6)$ in Voigt notation. The three kinds of distortions are: $(0 \quad 0 \quad 0 \quad \delta \quad \delta \quad \delta)$, $(\delta \quad \delta \quad 0 \quad 0 \quad 0 \quad 0)$, and $(\delta \quad \delta \quad \delta \quad 0 \quad 0 \quad 0)$. The distortions are chosen such that the resulting energy-strain fitting equations are linearly independent. Having obtained the elastic tensor elements, we compute elastic moduli as follows [8,9]:

$$B = \frac{1}{3}(C_{11} + 2C_{12}), \quad (6)$$

$$G = C_{44}, \quad (7)$$

$$E = \frac{9BG}{3B+G}, \quad (8)$$

$$\nu = \frac{3B-2G}{2(3B+G)}. \quad (9)$$

We note that the expressions for the bulk (B) and shear (G) moduli, Eqs. (6)–(7), are valid for cubic crystals, whereas the expressions for the Young's modulus (E) and Poisson's ratio (ν), Eqs. (8)–(9), are general for any crystal symmetry. [8,9] The fractional coordinates of each structure were optimized before applying distortions. $\delta$ was chosen to be ±0.01, ±0.02, and ±0.03, as suggested from a previous study [9].

**B. *NPT* MD method**

In the MD method, we calculate elastic properties from strain fluctuations via *NPT* ensemble MLFF MD simulation [10,43]. We first cast time-dependent lattice vectors **a**, **b**, and **c** into an upper-triangular form in the matrix **h**, satisfying the following conditions: **a** lies on the positive x axis, **b** is in the xy plane, with strictly positive y component and **c** may have any orientation with strictly positive z component [44].

$$\mathbf{h} = \begin{pmatrix} ||\mathbf{a}|| & ||\mathbf{b}||\cos\gamma & ||\boldsymbol{c}||\cos\beta \\ 0 & ||\mathbf{b}||\sin\gamma & (\mathbf{b}\cdot\mathbf{c} - b_x c_x)/b_y \\ 0 & 0 & \sqrt{||\boldsymbol{c}||^2 - c_x^2 - c_y^2} \end{pmatrix} \quad (10)$$

Here $\beta$ is the angle between **a** and **c**, and $\gamma$ is the angle between **a** and **b**. The strain tensor $\boldsymbol{\epsilon}$ can be calculated from **h**:

$$\boldsymbol{\epsilon} = \frac{1}{2}(\langle\mathbf{h}\rangle^{\mathrm{T},-1}\mathcal{G}\langle\mathbf{h}\rangle^{-1} - \mathbf{I}), \quad (11)$$

where $\mathcal{G} = \mathbf{h}^T\mathbf{h}$ is the metric matrix. The strain tensor $\boldsymbol{\epsilon}$ is then reduced to Voigt notation $\tilde{\boldsymbol{\epsilon}}$:

$$\tilde{\boldsymbol{\epsilon}} = \left(\epsilon_{xx}, \epsilon_{yy}, \epsilon_{zz}, \epsilon_{yz} + \epsilon_{zy}, \epsilon_{xz} + \epsilon_{zx}, \epsilon_{xy} + \epsilon_{yx}\right) \equiv (\epsilon_1, \epsilon_2, \epsilon_3, \epsilon_4, \epsilon_5, \epsilon_6). \quad (12)$$

The 6 by 6 symmetric elastic compliance matrix **S** and elastic stiffness matrix **C** can be calculated from the strain fluctuations [10,43,44],

$$S_{ij} = \frac{\langle V\rangle}{k_B T}\langle\Delta\epsilon_i\Delta\epsilon_j\rangle = \frac{\langle V\rangle}{k_B T}\left(\langle\epsilon_i\epsilon_j\rangle - \langle\epsilon_i\rangle\langle\epsilon_j\rangle\right), \quad (13)$$

$$\boldsymbol{C} = \boldsymbol{S}^{-1}. \quad (14)$$

Using the full elastic tensor, we then compute the elastic moduli under the Voigt-Reuss-Hill approximation [45,46]. Assuming uniform strain:

$$B_V = \frac{1}{9}[c_{11} + c_{22} + c_{33} + 2(c_{12} + c_{23} + c_{31})] \quad (15)$$

$$G_V = \frac{1}{15}[c_{11} + c_{22} + c_{33} - (c_{12} + c_{23} + c_{31}) + 3(c_{44} + c_{55} + c_{66})]. \quad (16)$$

Assuming uniform stress:

$$B_R = \frac{1}{s_{11}+s_{22}+s_{33}+2(s_{12}+s_{23}+s_{31})} \quad (17)$$

$$G_R = \frac{15}{4(s_{11}+s_{22}+s_{33})-4(s_{12}+s_{23}+s_{31})+3(s_{44}+s_{55}+s_{66})} \quad (18)$$

We elect to take the arithmetic average of the two as suggested [45,46]:

$$B = (B_V + B_R)/2 \quad (19)$$

$$G = (G_V + G_R)/2. \quad (20)$$

The Young's modulus and Poisson's ratio were computed the same way as in the static case. To obtain statistically meaningful results, we used standard error propagation to evaluate the standard deviations of elastic constants and moduli [10]. In the following equations, we ignore all the cross-correlated terms [47]:

$$Var(S_{ij}) = \left(\frac{S_{ij}}{\langle V\rangle}\right)^2 Var(V) + \left(\frac{\langle V\rangle}{k_B T}\right)^2 \left[Var(\epsilon_i\epsilon_j) + \langle\epsilon_j\rangle^2 Var(\epsilon_i) + \langle\epsilon_i\rangle^2 Var(\epsilon_j)\right]. \quad (21)$$

Note that $d(CS) = dI = 0$, so $dC = -CdSC$. Thus $dC_{ij} = -\sum_{k=1}^{n}\sum_{l=1}^{n} C_{ik} dS_{kl} C_{lj}$ and $\frac{\partial C_{ij}}{\partial S_{kl}} = -C_{ik}C_{lj}$. This leads to:

$$Var(C_{ij}) = \sum_{k=1}^{n}\sum_{l=1}^{n} C_{ik}^2 Var(S_{kl}) C_{lj}^2 \quad . (22)$$

With $Var(S_{ij})$ and $Var(C_{ij})$, it is straightforward to evaluate the variances of elastic moduli. The data and the code to calculate elastic properties and their variances from VASP output can be found on GitHub [48].

## III. RESULTS AND DISCUSSIONS

### A. Elastic properties from static methods

Being simple and computationally efficient, static methods to predict elastic properties have been widely used in crystalline SSEs [8,9,49], amorphous borides [50], and amorphous carbon materials [51]. Here, as a first step, we apply these static methods to the amorphous $LiTaCl_6$ shown in Fig. 1(a). For the unrelaxed static method, we plot the energy of a randomly chosen structure of amorphous $LiTaCl_6$ (taken from the NVT AIMD at 300 K shown in Fig. S1) with the internal coordinates fixed against the uniform strain for three different distortions. As can be seen in Fig. 1(b), for a 3% strain, the energy increases to 2 to 10 eV per cell (roughly, 80 to 400 meV/atom) above the equilibrium value. After considering ionic relaxation, the energy changes are about one order of magnitude smaller [Fig. 1(c)], indicating that the amorphous structure has many degrees of freedom to lower energy upon applied strain.

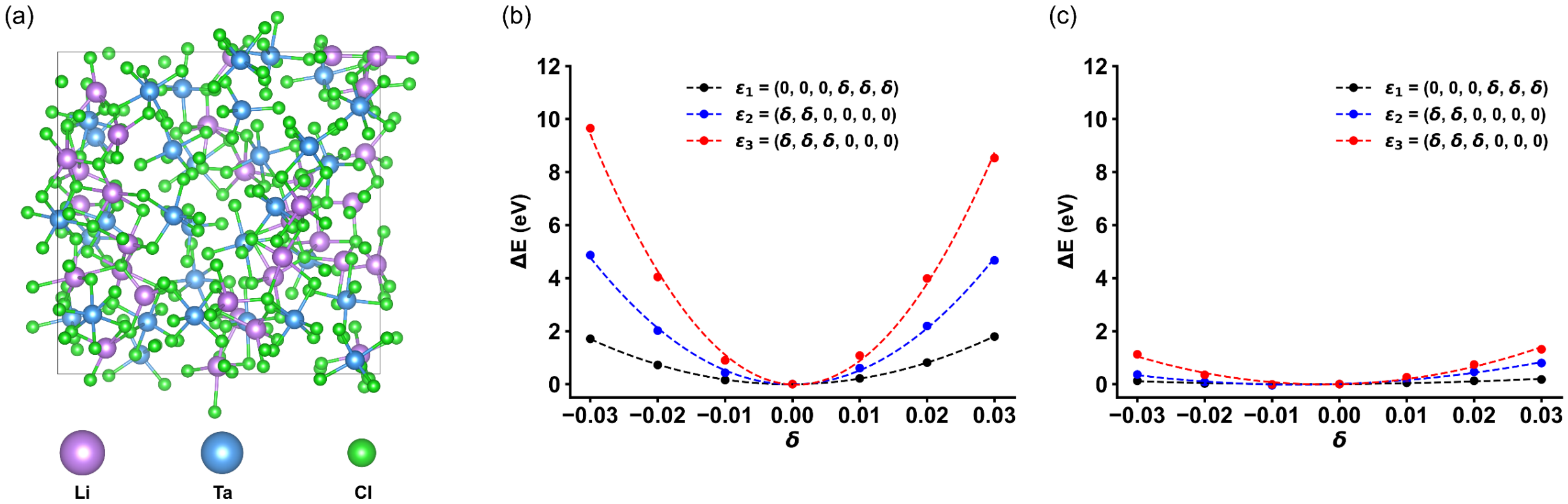

**FIG. 1.** (a) One snapshot from the AIMD trajectory of the amorphous $LiTaCl_6$. There are 32 $LiTaCl_6$ units in the cubic cell of 19.30 Å in edge length. (b) Unrelaxed static energy (E) vs strain (δ) for three different distortions ($\boldsymbol{\varepsilon}$) on a randomly chosen structure of the amorphous $LiTaCl_6$. (c) Relaxed static energy vs strain for three different distortions on a randomly chosen structure of the amorphous $LiTaCl_6$.

From the fitting lines in Fig. 1(b)-1(c), we extracted the Young's moduli using Eqs. (5)-(8) from both the unrelaxed and relaxed static approaches for this one random structure. We repeated the process for 9 more different random structures and averaged the values; these 10 random structures were sufficient to obtain statistically significant differences between the unrelaxed and relaxed static approaches. The computed Young's moduli are 73.79 ± 1.70 GPa from the unrelaxed static approach and 7.91 ± 1.59 GPa from the relaxed static approach, while the experimental value is 2.91 ± 0.32 GPa [14] (Fig. 2). The unrelaxed static approach leads to a Young's modulus 25 times the experimental value and is qualitatively incorrect. Although the unrelaxed static approach has been shown to overestimate the bulk modulus of crystalline superionic t-LGPS by about 100% [10], the approximately 25-fold overestimation of Young's modulus in amorphous $LiTaCl_6$ demonstrates that this approach, while suited for crystalline systems with limited degrees of freedom to relax, becomes qualitatively invalid for the amorphous system. The relaxed static approach, despite being in the same order of magnitude as the experimental value, still overestimates it by more than 170%. Even when we increase the system size to a cubic 2×2×2 supercell (2048 atoms, matching the cell later employed in the MLFF–MD calculations), the relaxed static method still significantly overestimates the Young's modulus by more than 80% (Fig. S5; Table S2 [24]). This overestimation suggests that just relaxing the ionic positions to a local minimum cannot sufficiently sample all the energy configurations accessible to the amorphous structure subjected to a distortion. This reflects the limitation of static approaches, whose harmonic and affine deformation assumptions break down in amorphous superionic materials, where the elastic response can only be captured through dynamical sampling of the relevant configurational ensemble during deformation.

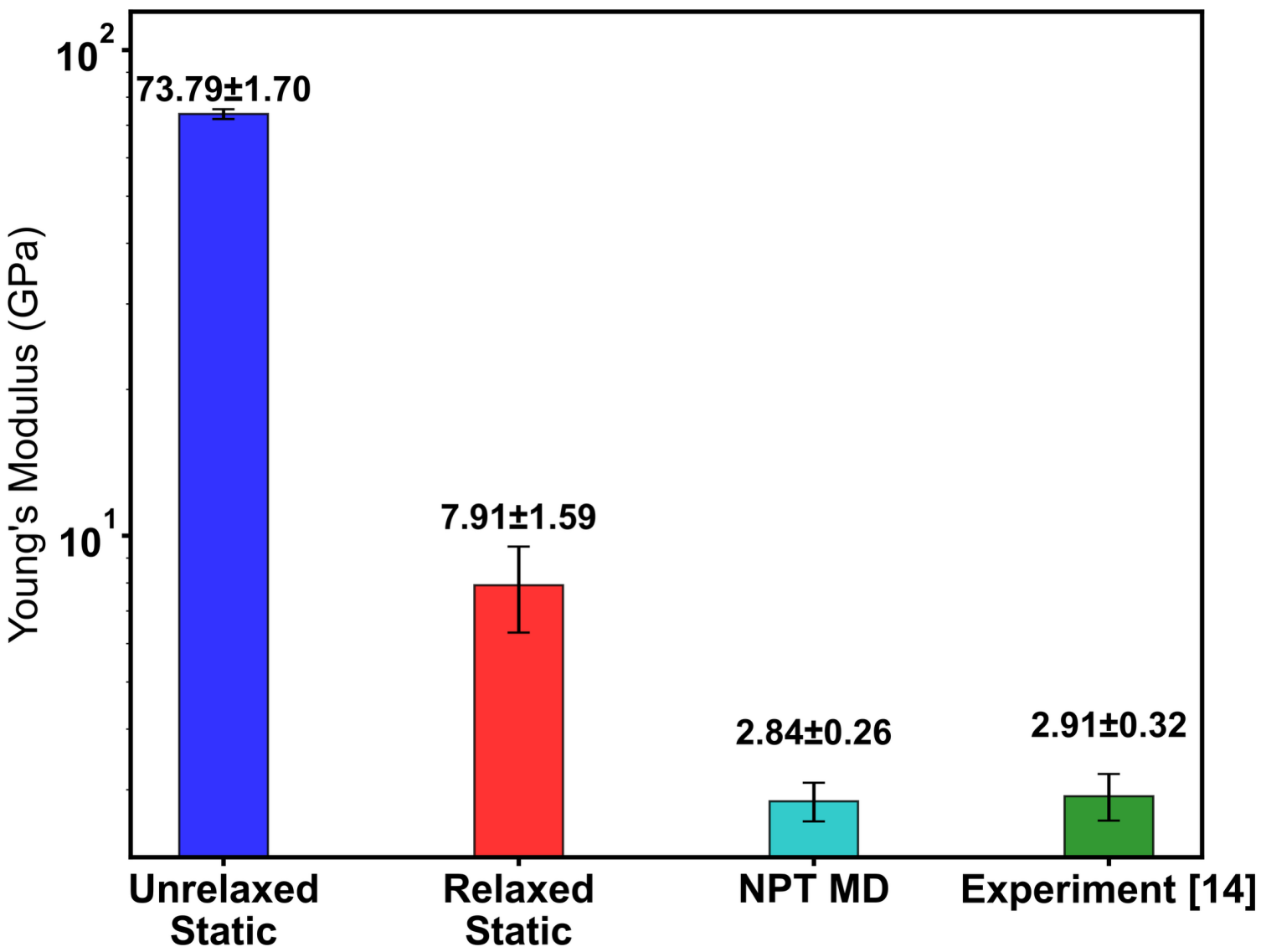

**FIG. 2.** Comparison of the computed Young's moduli of amorphous $LiTaCl_6$ from three different methods against the experimental value [14].

## B. Elastic properties from the *NPT* molecular dynamics method

To more accurately predict elastic properties of amorphous $LiTaCl_6$, we need to sample more configurations by using a larger simulation cell and longer sampling time. Therefore, we employed the MLFF MD in the *NPT* ensemble to simulate a cubic supercell containing 2048 atoms [Fig. 3(a)] for 12.0 ns with the first 8.0 ns for equilibration and the last 4.0 ns for analysis (Fig. S6 [24]). From the trajectory, we extracted the moduli from the fluctuations of the strain as described in Eqs. (10-20). To obtain an estimate of statistical errors of the elastic constants and moduli, we carried out a block analysis of the relative standard deviations, following a previous study [10,52]. As shown in Fig. 3(b), the optimal block size is determined by the onset of the plateau [52,53], which is 7500 data-in-block and equivalent to 150 ps. The convergence of the elastic moduli is shown in Figure 4, where the mean values and standard deviations are reported (cf. [10]). All elastic moduli converge after 3.0 ns. However, the standard deviations remain large, and the fact that these relative standard deviations are larger than those obtained using the same method for crystalline SSEs [10] suggests that this may be due to the amorphous nature of the material, which leads to greater inherent structural fluctuations and heterogeneity. Nevertheless, the converged value of the Young's modulus (3.09 ± 0.55 GPa) is in very good agreement with the experimental value of 2.91 ± 0.32 GPa [14] (Fig. 2). To tighten the error bounds, we ran two additional trajectories (Table S3; Figs. S7 and S8 [24]) and obtained an average Young's modulus of 2.84 ± 0.26 GPa. Time-dependent $\bar{q}_6$ distribution and $\langle\bar{q}_6\rangle$ show that the three trajectories are clearly distinct (Fig. S9 [24]); the $\langle\bar{q}_6\rangle$ values become more stabilized in the last 4 ns, with traj 1 being slightly more amorphous ($\langle\bar{q}_6\rangle = 0.150 \pm 0.004$) than traj 2 ($\langle\bar{q}_6\rangle = 0.155 \pm 0.004$), which is in turn more amorphous than traj 3 ($\langle\bar{q}_6\rangle = 0.170 \pm 0.004$). The final, converged values of the elastic moduli of amorphous $LiTaCl_6$ predicted from MLFF-MD are shown in Table I, together with their standard deviations. Overall, amorphous $LiTaCl_6$ behaves elastically like a soft polymer or gel such as polypropylene, as the comparison in Table I shows.

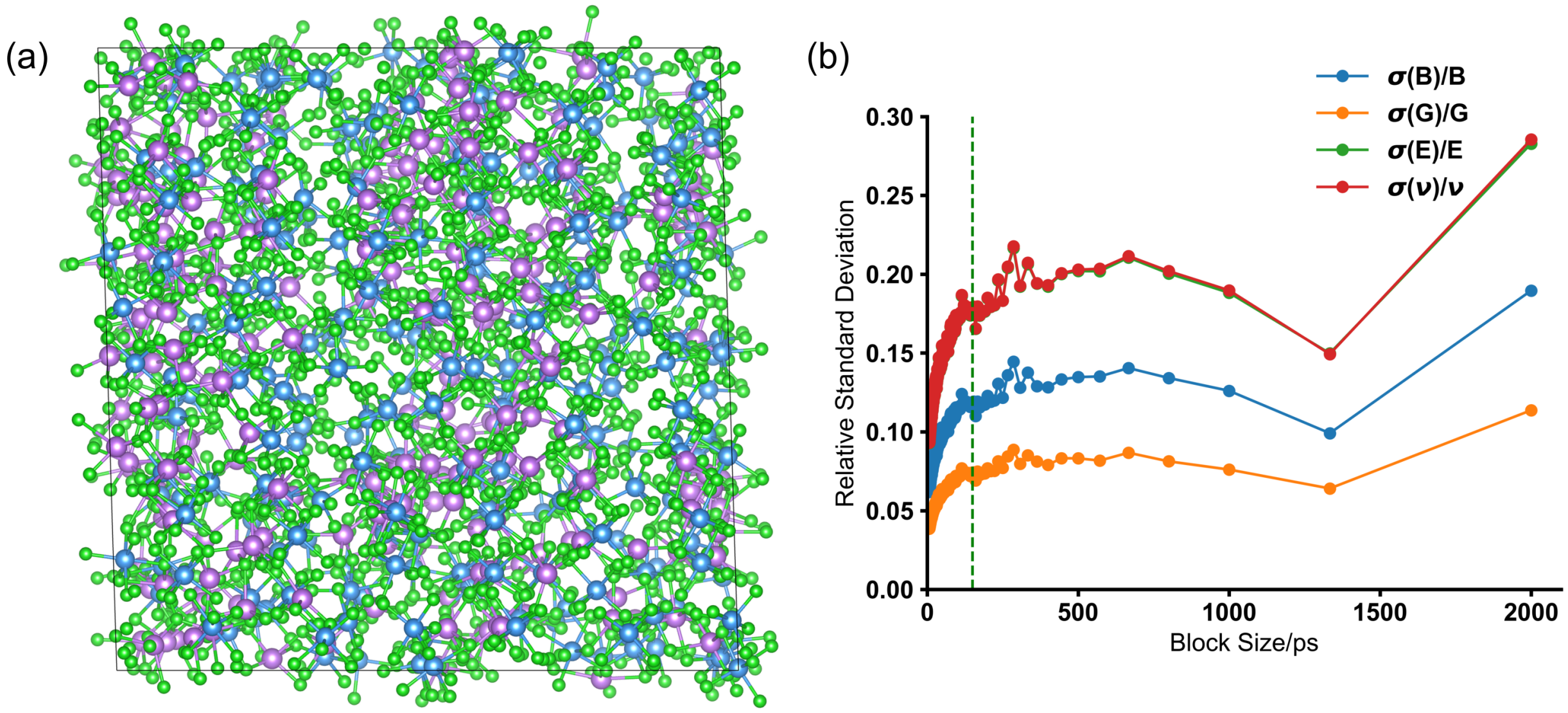

**FIG. 3.** (a) Snapshot of the amorphous $LiTaCl_6$ in a supercell of 38.6 Å in edge length (256 $LiTaCl_6$ units, 2048 atoms) used in the *NPT* MLFF-MD simulation. (b) Relative standard deviation of elastic properties vs block size: B, bulk modulus; G, shear modulus; E, Young's modulus; ν: Poisson's ratio. The dashed green line indicates that the relative deviations all reach a plateau at a block size of 150 ps.

To test the transferability of the trained MLFF and the sensitivity of the simulated elastic properties to other possible amorphous structures of $LiTaCl_6$, we tested our MLFF on a more amorphous initial structure of $LiTaCl_6$ with a significantly lower initial $\langle\bar{q}_6\rangle$ of 0.135, generated from purely geometric random packing (see Table S4 notes [24] for details). After 12 ns of simulation, $\langle\bar{q}_6\rangle$ of the random-packed structure stabilized around 0.148 ± 0.003, which remains more amorphous than the three trajectories above using the AIMD/MLFF training initial structure;

more importantly, its simulated mechanical properties (Table S4 [24]) agree well with those from averaging the three trajectories (Table I; Table S3 [24]). So we conclude that our MLFF can indeed generalize well to the most common cases of amorphous structures of $LiTaCl_6$ where the $[TaCl_6]^-$ octahedra are randomly packed with $Li^+$ ions occupying interstitial sites. In the rare cases of fused/networked $[TaCl_6]^-$ octahedra with free $Cl^-$ ions, further sampling and training would be needed.

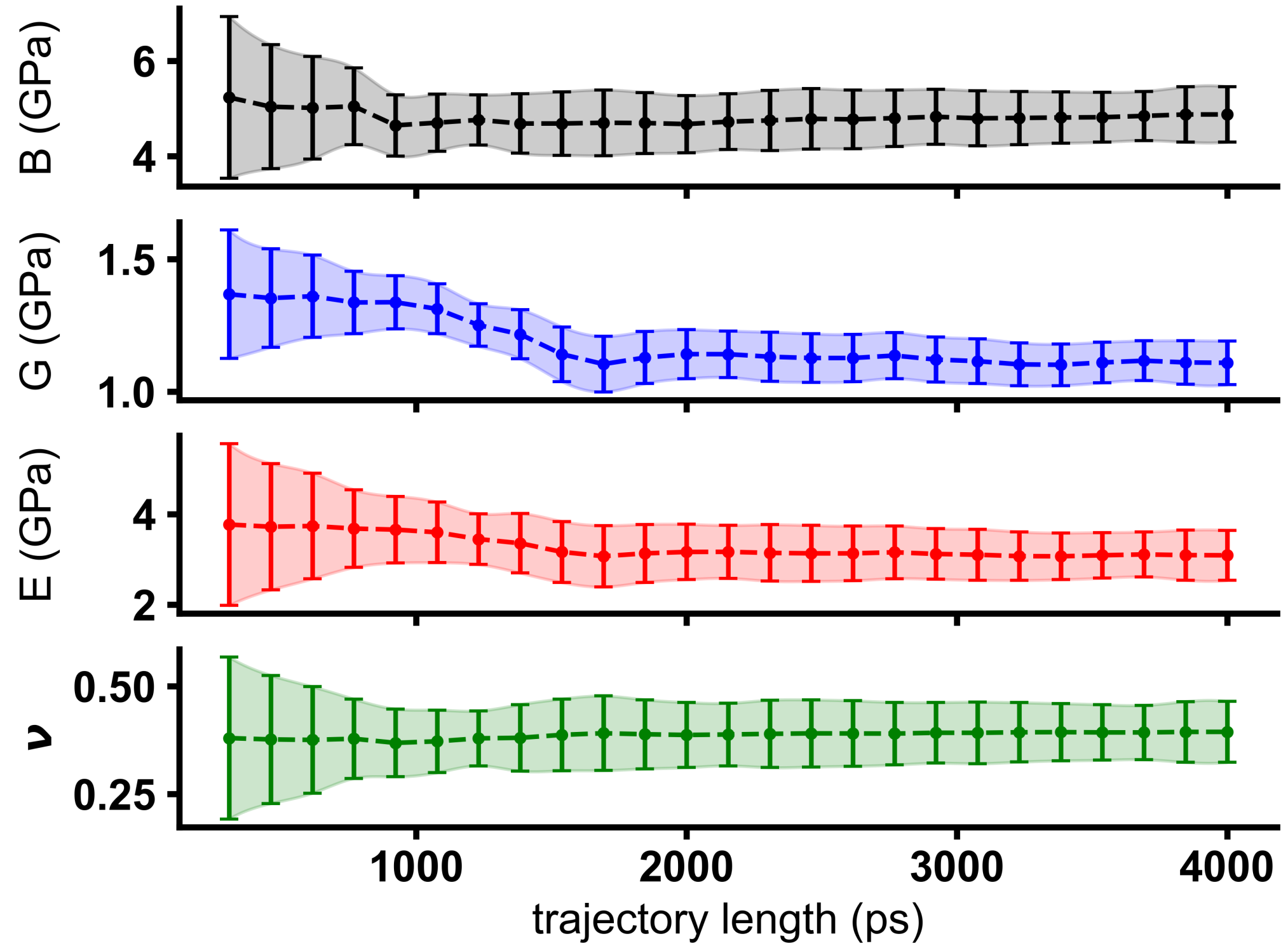

**FIG. 4.** Convergence of elastic properties with the trajectory length from *NPT* MLFF-MD simulation of the amorphous $LiTaCl_6$ at 300 K and 1 bar.

TABLE I. Elastic properties of amorphous $LiTaCl_6$ predicted from MLFF-MD in this work in comparison with those of polypropylene [18].

| **Elastic properties** | **$LiTaCl_6$** | **Polypropylene** [18] |
|---|---|---|
| **Young's modulus (E)/GPa** | 2.84 ± 0.26 | 1.847 |
| **Shear modulus (G)/GPa** | 1.02 ± 0.04 | 0.650 |
| **Bulk modulus (B)/GPa** | 4.44 ± 0.26 | 3.80 |
| **Poisson's ratio ($\nu$)** | 0.39 ± 0.04 | 0.42 |

### C. Comparison with other superionic conductors

We compare the novel amorphous $LiTaCl_6$ electrolyte with other established superionic conductors and discuss its implications for practical applications. Fig. 5 shows the elastic properties of the $LiTaCl_6$ with bcc Li and some selected SSEs. In contrast with the other popular SSEs which are all stiffer than Li, the amorphous $LiTaCl_6$ is more compliant. The softness of the material, combined with high ductility (Fig. 6 (a)) inferred from elastic properties [19,20,23] indicates ease of processing of the material. However, the calculated shear modulus of amorphous $LiTaCl_6$ is only 1.02 GPa, which is far smaller than twice the shear modulus of bcc Li that is believed to be the minimum value necessary to suppress Li dendrite initiation [6]. Therefore, the anode compatibility could be a challenge for $LiTaCl_6$ given its low shear modulus and the potential instability of $Ta^{5+}$ against reduction by Li metal. The high Poisson's ratio of $LiTaCl_6$ is comparable to Li metal, halide, and sulfide SSEs, while being higher than oxyhalide and oxide electrolytes (Fig. 6 (b)). It indicates that $LiTaCl_6$ has greater lateral compliance, which can facilitate conformal interfacial contact with electrodes under stack pressure.

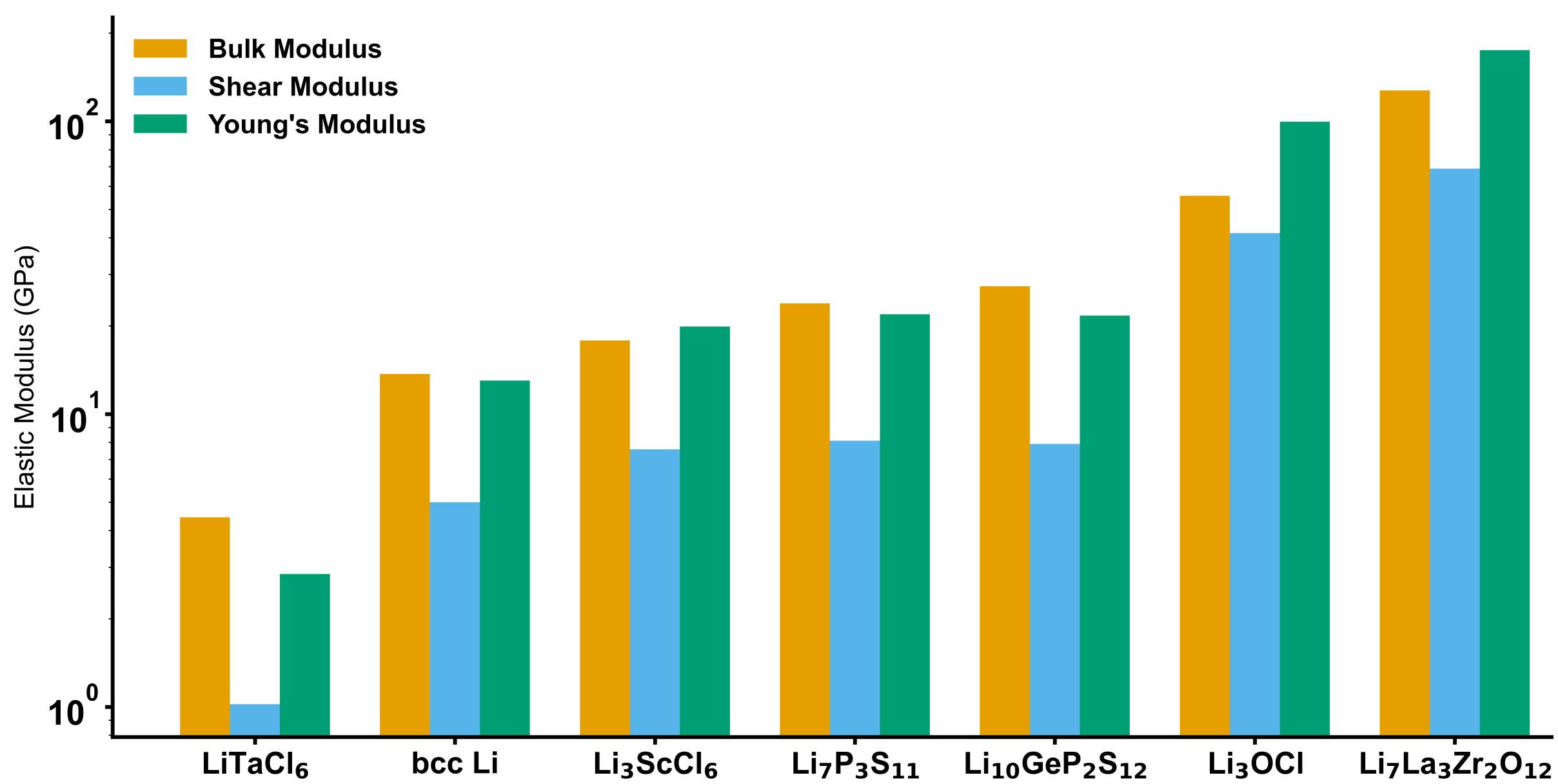

**FIG. 5.** Comparison of the MLFF-MD simulated elastic moduli of amorphous $LiTaCl_6$ of the present work with DFT-computed values of bcc Li and selected SSEs from the literature. The bcc Li data are taken from Ref. [9], the $Li_3ScCl_6$ data from Ref. [54], and the rest from Ref. [8].

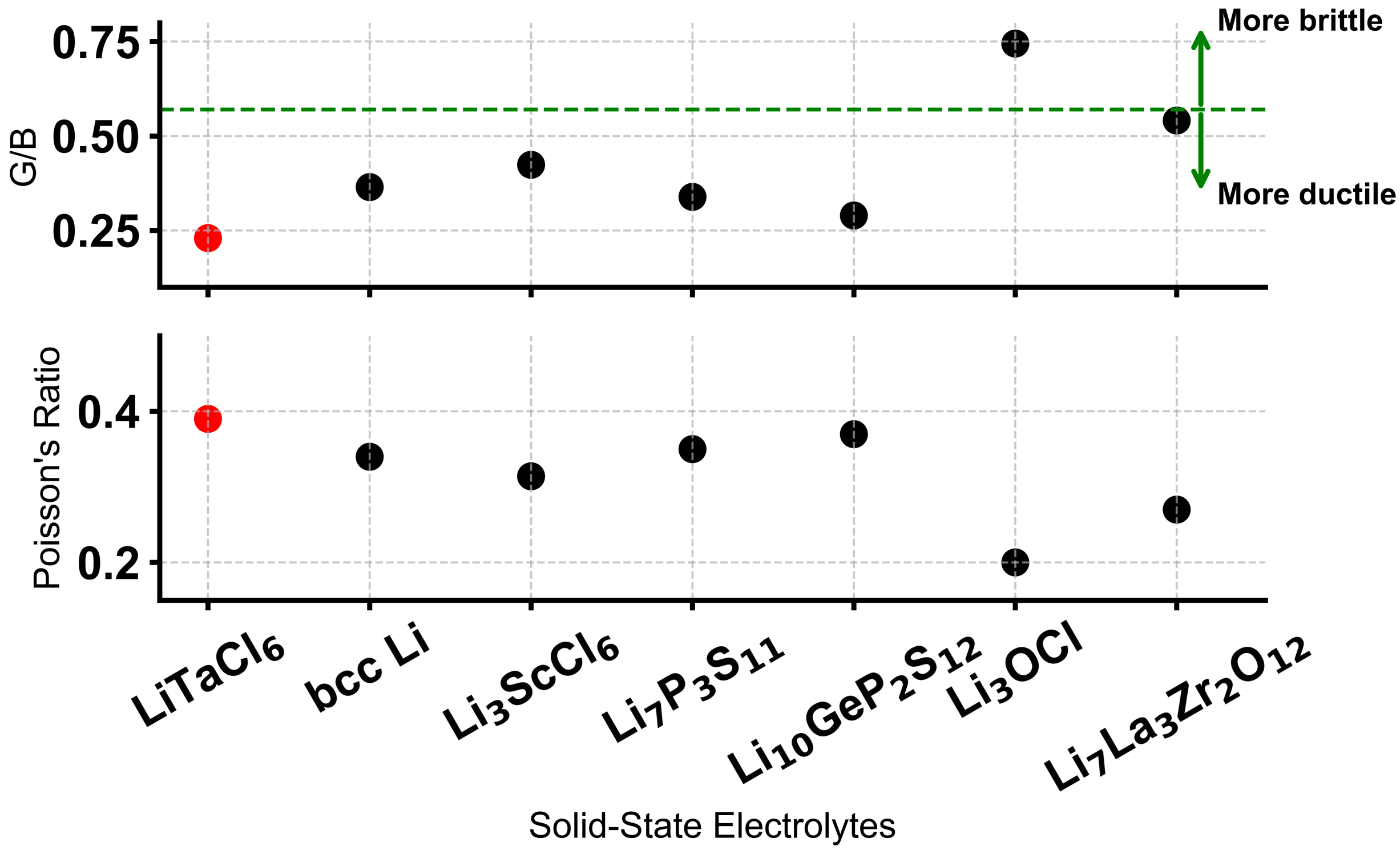

**FIG. 6.** Comparison of the MLFF-MD simulated Pugh ratio or G/B (top) and Poisson's ratio (bottom) of the amorphous $LiTaCl_6$ of the present work with DFT-computed values of selected SSEs from the literature. The bcc Li data are taken from Ref. [9], the $Li_3ScCl_6$ data from Ref. [54], and the rest from Ref. [8].

## IV. CONCLUSIONS

In this study, we present a comprehensive evaluation of the elastic properties of the amorphous superionic $LiTaCl_6$ using three computational methods. We found that MD simulations with MLFFs of a 2000-atom supercell of amorphous $LiTaCl_6$ yielded a Young's modulus of 2.84 ± 0.26 GPa—remarkably close to the experimental value of 2.91 ± 0.32 GPa [14]. MLFF–MD simulations also predicted a bulk modulus of 4.44 GPa, a shear modulus of 1.02 GPa, and a Poisson's ratio of 0.39, indicating that this material behaves elastically like a soft polymer such as polypropylene. Our work highlights the distinctive elastic behavior of amorphous superionic materials and underscores the importance of dynamical sampling in capturing two key features: anharmonic effects arising from their superionic nature at room temperature, and non-affine deformation under applied strain characteristic of amorphous materials. By leveraging nanosecond-scale molecular dynamics simulations enabled by machine-learning force fields, we present an accurate and efficient approach for evaluating the elastic properties of amorphous superionic conductors.

**Acknowledgements**

This research was funded by the U.S. Department of Energy, Office of Science, Office of Basic Energy Sciences Award DE-SC0023408. This research used resources of the National Energy Research Scientific Computing Center, a DOE Office of Science User Facility supported by the Office of Science of the U.S. Department of Energy under contract no. DE-AC02-05CH11231.

**Data Availability**

The data that support the findings of this study are available from GitHub [48].

*Supplemental Material*

# Elastic Properties of Amorphous $LiTaCl_6$ Solid-State Electrolyte

Xiaolin Liu[1] and De-en Jiang[1, *]

[1] Department of Chemical and Biomolecular Engineering, Vanderbilt University, Nashville, Tennessee 37235, USA

*E-mail: de-en.jiang@vanderbilt.edu

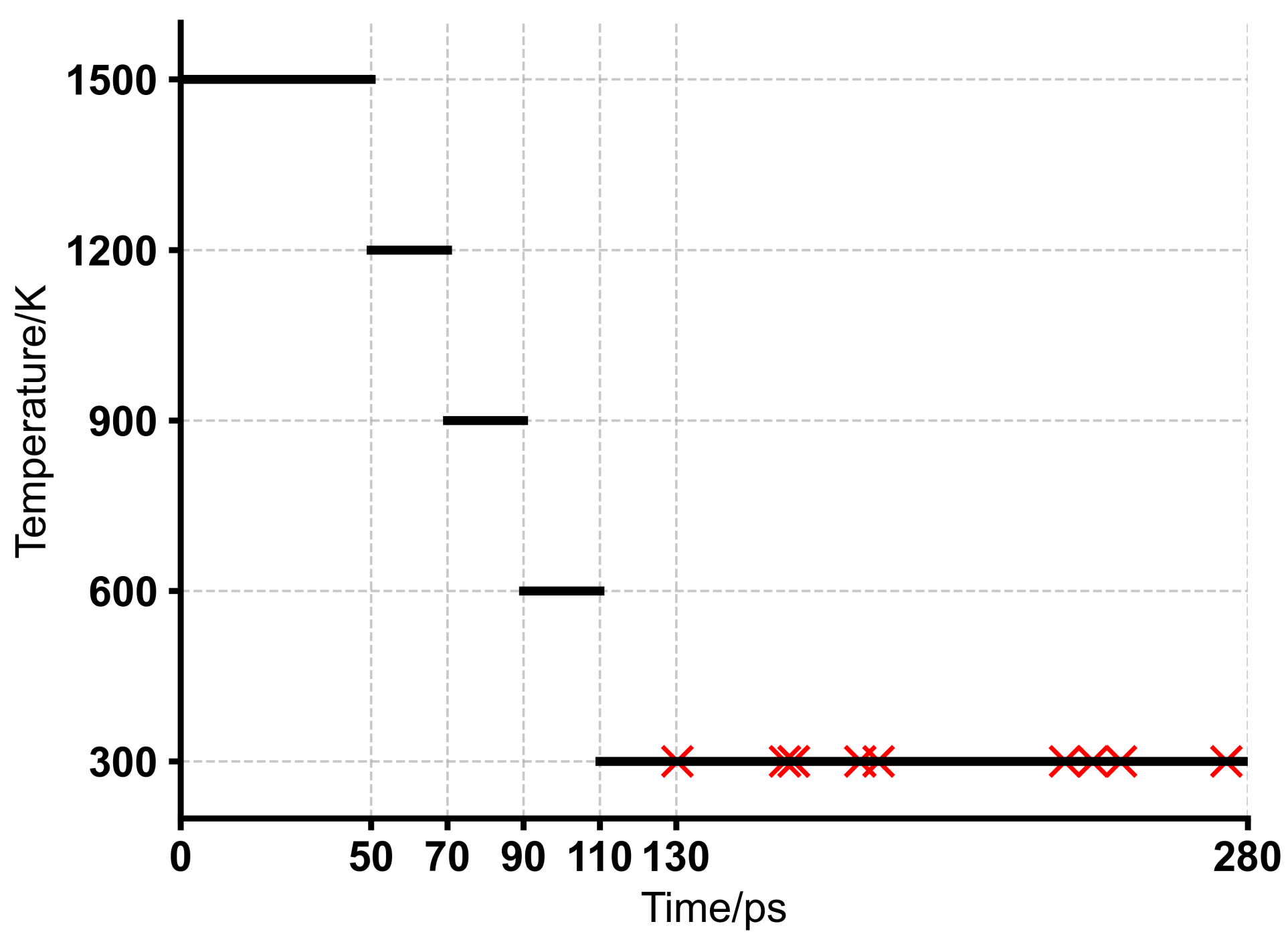

FIG. S1. Temperature schedule used to generate the amorphous $LiTaCl_6$ structure. The horizontal black segments indicate the NVT–AIMD simulations used for melting at 1500 K, stepwise quenching to 300 K, and subsequent production sampling at 300 K. Equilibrium volumes at 1500 K and 300 K were verified separately using short NPT simulations and, at 1500 K, by evaluating the internal energy as a function of a uniform lattice-scaling factor (not shown). The final 300 K trajectory was propagated entirely in the NVT ensemble, from which ten random snapshots (red crosses) were selected for the static elastic calculations.

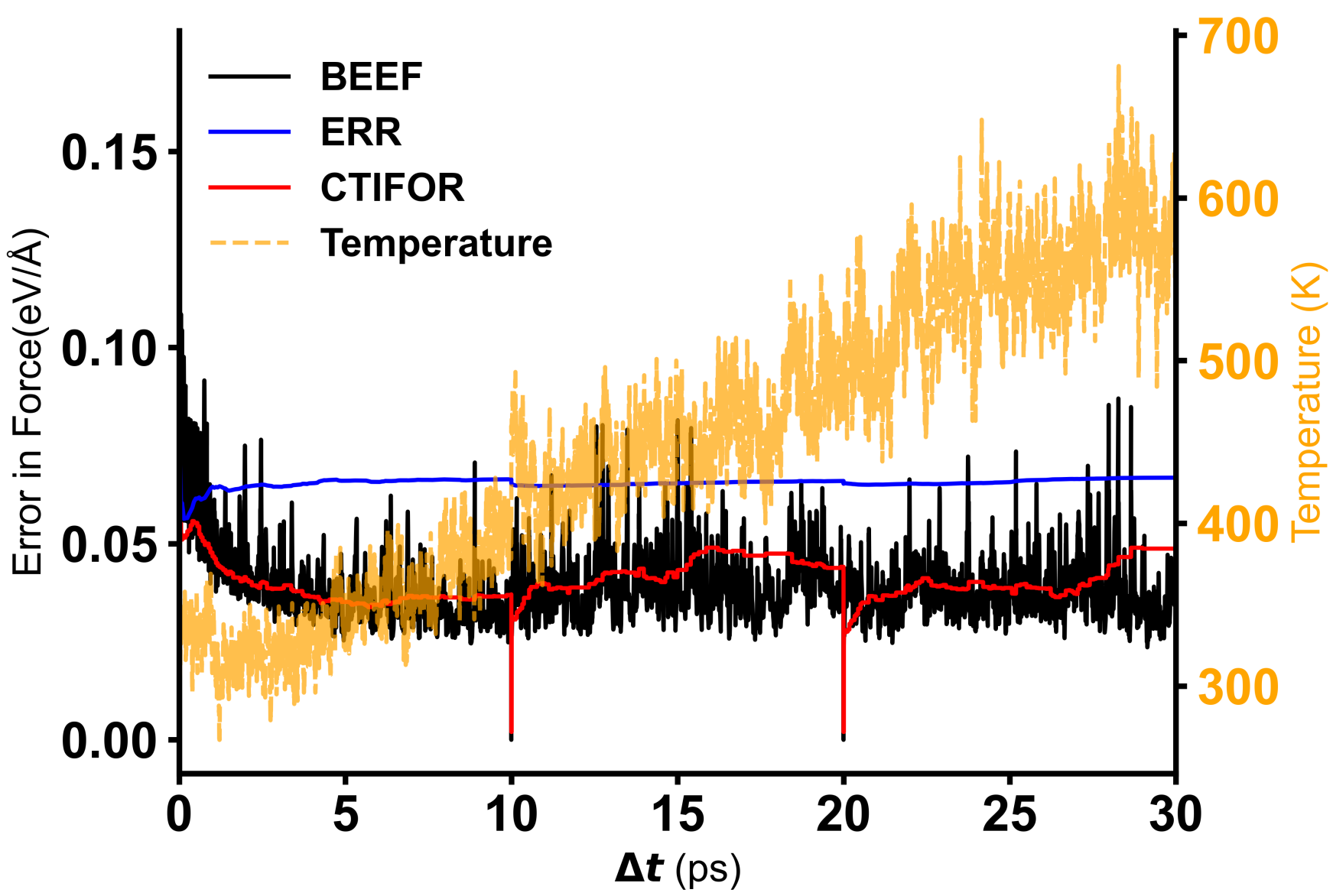

FIG. S2. Evolution of force errors and temperature during a 30-ps temperature ramp used for machine-learning force field training. The temperature (right y-axis, orange dashed line) increases from 300 K to 600 K in three 10-ps linear ramps: 300–400 K, 400–500 K, and 500–600 K. The left y-axis shows the force prediction error (ERR, blue), Bayesian error estimate (BEEF, black), and error threshold (CTIFOR, red) used to trigger ab initio evaluations. The sudden drops in CTIFOR at 10 ps and 20 ps correspond to restarts at each ramp stage.

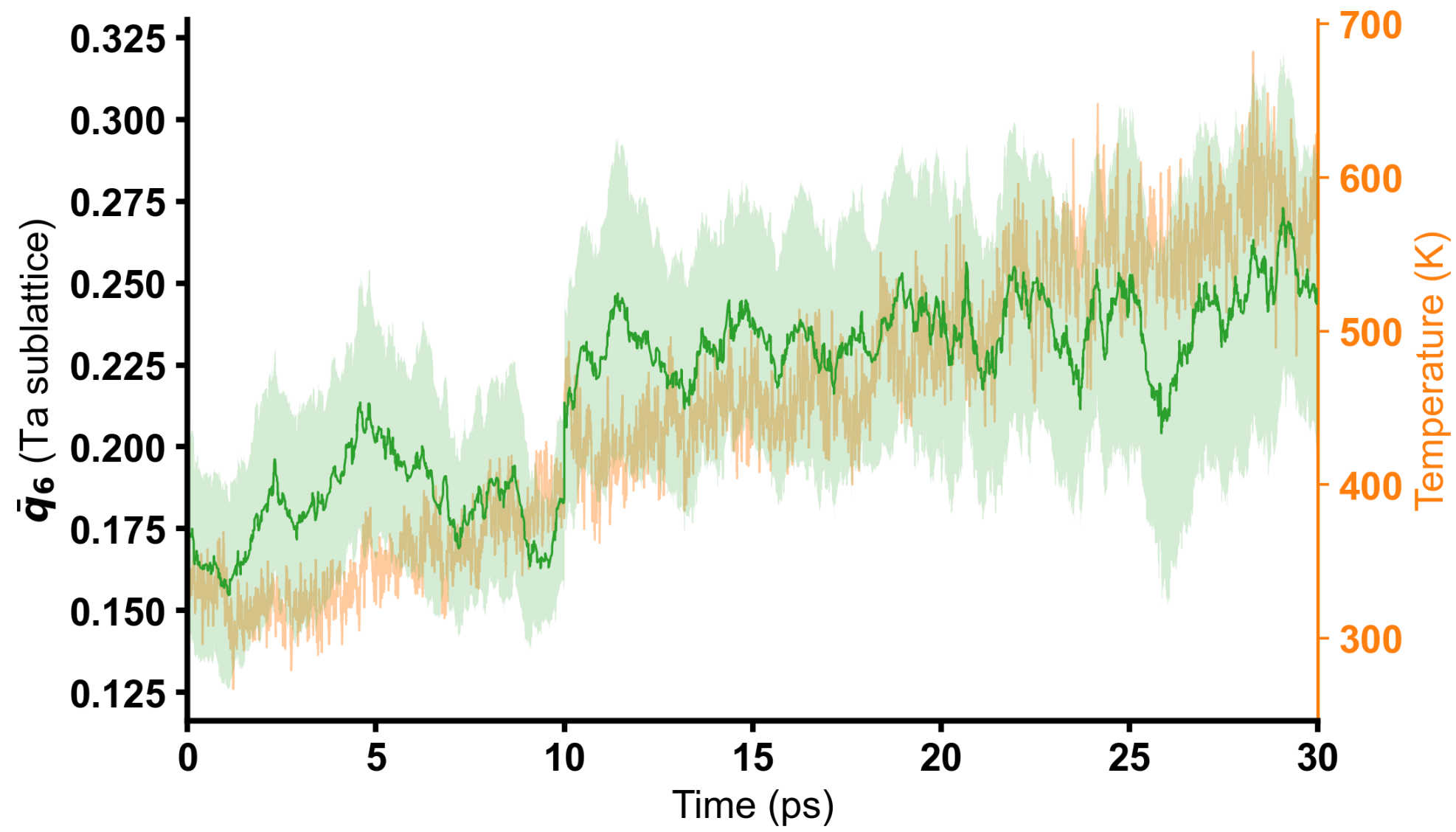

FIG. S3. The Lechner–Dellago neighbor-averaged bond-orientational order ($\bar{q}_6$; lighter green traces) of the Ta sublattice and its mean ($\langle\bar{q}_6\rangle$; darker green traces) for the 30-ps AIMD/MLFF training structures for temperature from 300 K to 600K (lighter orange traces).

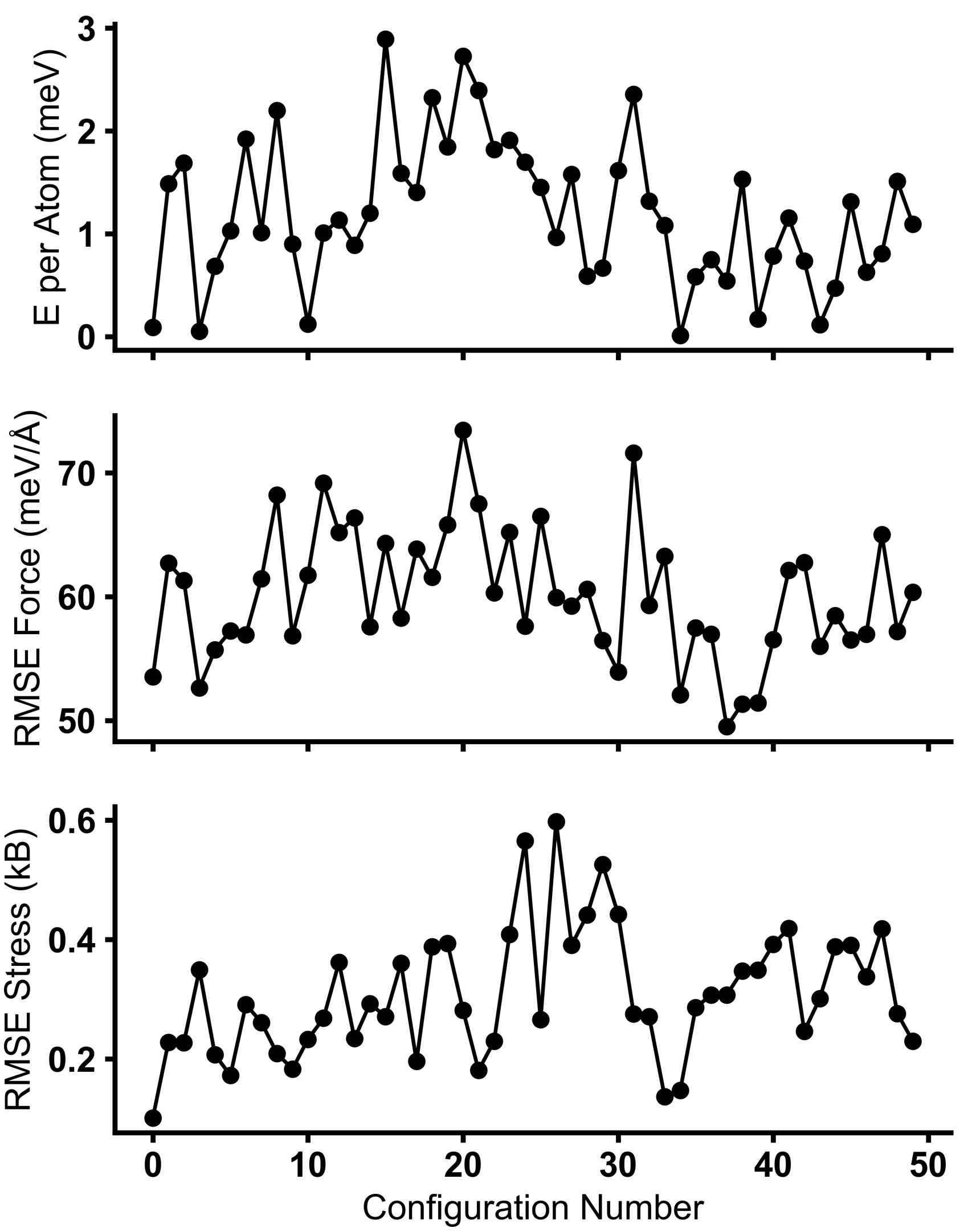

FIG. S4. The test errors are obtained from 50 snapshots at every 1ps from a 50 ps MLFF simulation at 300 K.

Table S1. Comparison of training and test errors. The training errors are obtained from the refit procedure implemented in the VASP software, and the test errors are from data plotted in Fig. S4.

| Errors | Training | Test |
|---|---|---|
| **E (meV/atom)** | 0.7 | 1.2 |
| **RMSE Force (meV/Å)** | 64.2 | 60.1 |
| **RMSE Stress (kB)** | 0.3 | 0.3 |

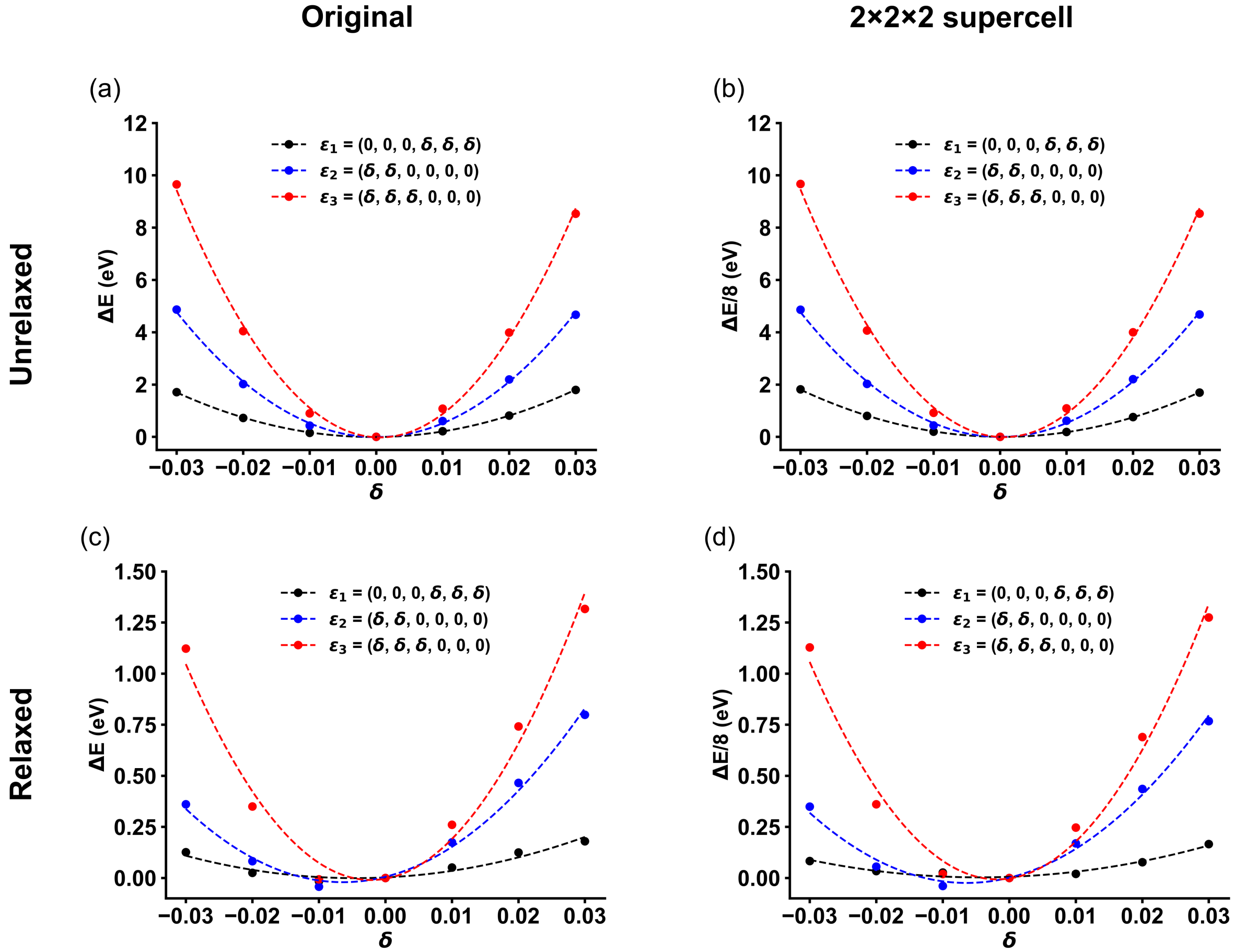

FIG. S5. Energy–strain relations for three independent lattice distortions computed for amorphous $LiTaCl_6$ using static methods. Panels (a) and (b) show unrelaxed static calculations for the original cubic cell (256 atoms) and a 2×2×2 supercell (2048 atoms), respectively. Panels (c) and (d) show the corresponding relaxed static calculations. For visualization purposes, the energies of the supercell calculations are scaled by a factor of 1/8 to facilitate direct comparison; however, the elastic constants were obtained from fits to the unscaled energies.

Table S2. Elastic properties of amorphous $LiTaCl_6$ calculated using unrelaxed and relaxed static methods for the small (256 atoms) and large (2048 atoms) cell sizes. For comparison, the experimental Young's modulus is 2.91 ± 0.32 GPa.[a]

| **Elastic Properties** | **Unrelaxed** | | **Relaxed** | |
|---|---|---|---|---|
| | Small cell | Large cell | Small cell | Large cell |
| **Young's modulus (E)/GPa** | 73.02 | 72.88 | 6.71 | 5.33 |
| **Shear modulus (G)/GPa** | 29.04 | 28.97 | 2.51 | 1.95 |
| **Bulk modulus (B)/GPa** | 50.11 | 50.13 | 6.74 | 6.61 |
| **Poisson's ratio ($\nu$)** | 0.26 | 0.26 | 0.33 | 0.37 |

[a] Note that the values reported in this table for a single representative structure.

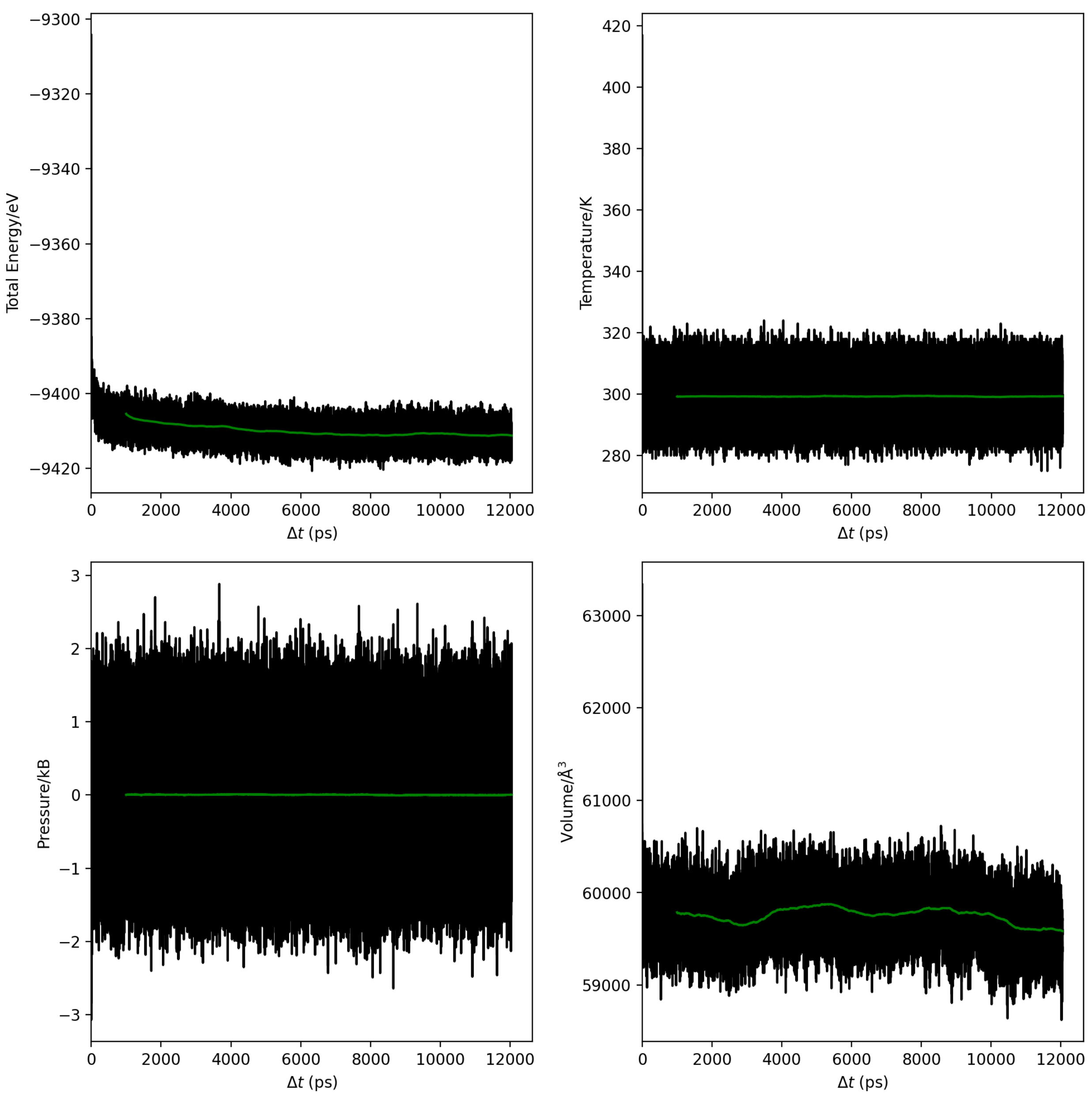

FIG. S6. The total energy, temperature, pressure, and volume during a 12.0 ns NPT MD simulation. The green curve is the moving average over 1 ns.

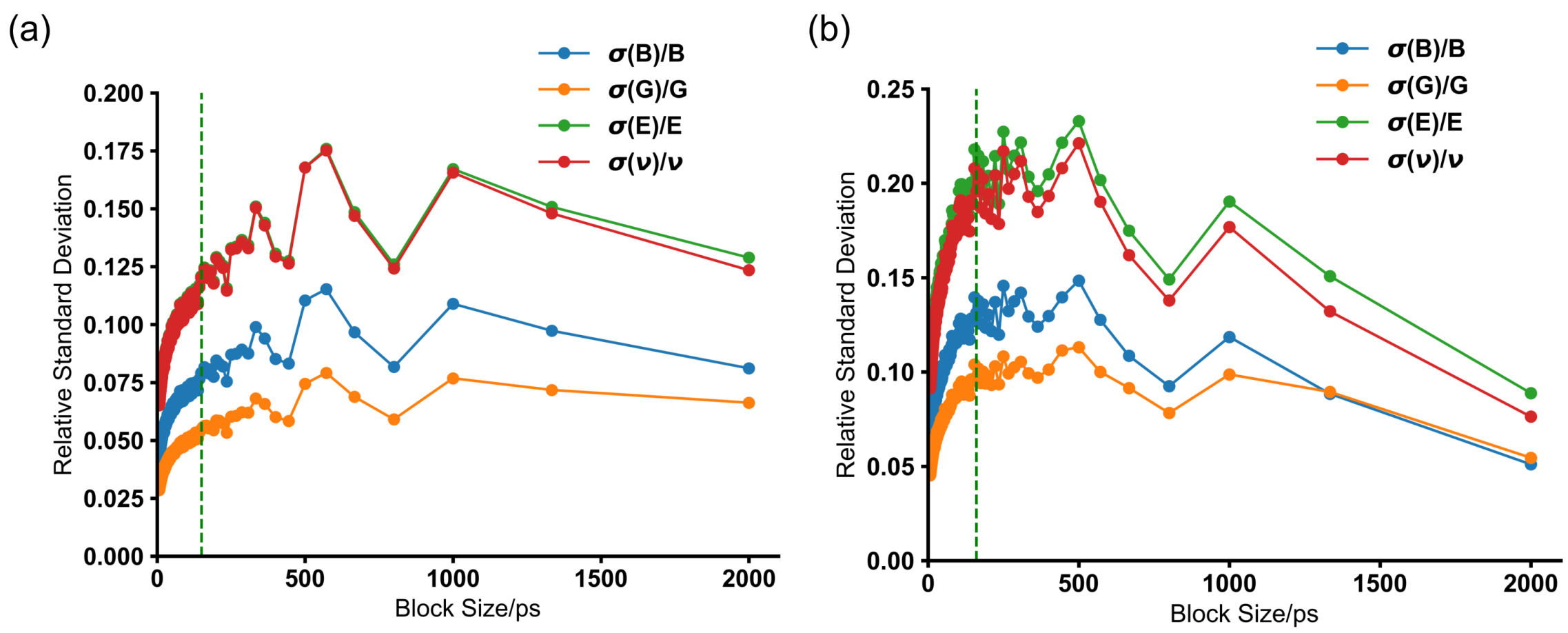

FIG. S7. Relative standard deviations of elastic properties versus block size for trajectories 2 (a) and 3 (b). The dashed green lines indicate that the relative deviations plateau at a block size of 150 ps (a) and 160 ps (b).

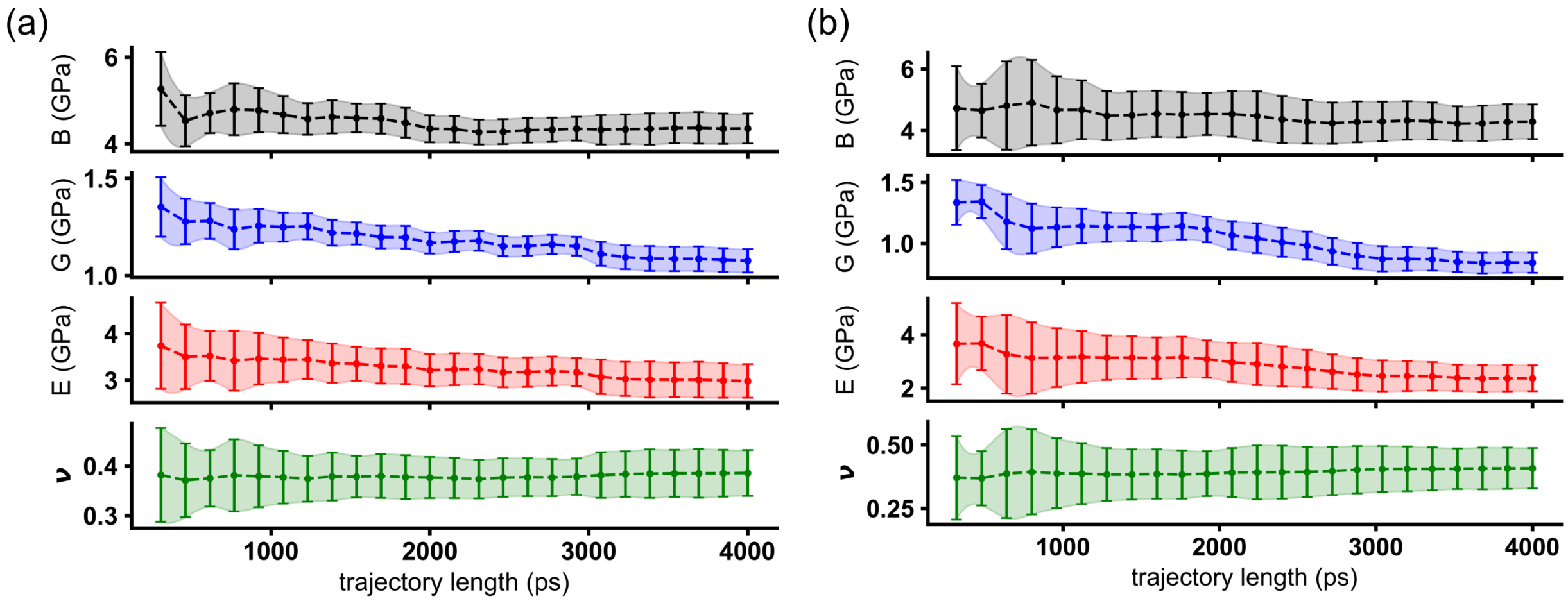

FIG. S8. Convergence of elastic properties with increasing trajectory length in NPT MLFF–MD simulations for trajectories 2 (a) and 3 (b).

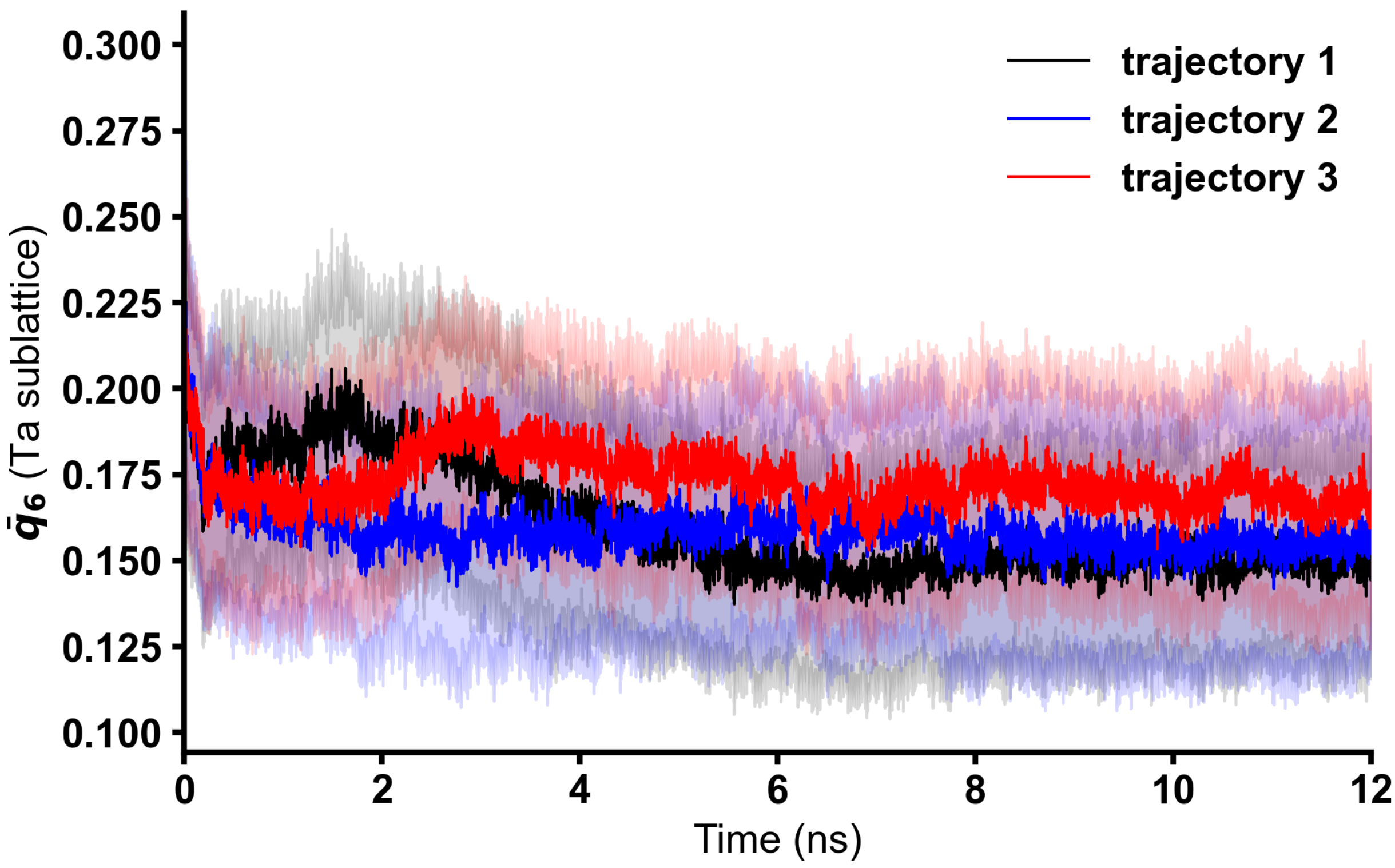

FIG. S9. The time evolution of the Lechner–Dellago neighbor-averaged bond-orientational order ($\bar{q}_6$; lighter traces) of the Ta sublattice and its mean ($\langle\bar{q}_6\rangle$; darker traces) for the three 12-ns NPT trajectories of $LiTaCl_6$ at 300 K; the three trajectories started with the same initial structure at the end of the on-the-fly MLFF training but with different initial random velocities.

Table S3. Raw elastic properties calculated from the three trajectories. The final values in TABLE I are obtained using the inverse variance weighting scheme.

| Elastic Properties | Trajectory Number | | |
|---|---|---|---|
| | 1 | 2 | 3 |
| **Young's modulus (E)/GPa** | 3.09 ± 0.55 | 2.98 ± 0.36 | 2.37 ± 0.49 |
| **Shear modulus (G)/GPa** | 1.11 ± 0.08 | 1.08 ± 0.06 | 0.84 ± 0.08 |
| **Bulk modulus (B)/GPa** | 4.87 ± 0.58 | 4.35 ± 0.34 | 4.28 ± 0.56 |
| **Poisson's ratio ($\nu$)** | 0.39 ± 0.07 | 0.39 ± 0.05 | 0.41 ± 0.08 |

Table S4. MLFF-MD-simulated elastic properties of amorphous $LiTaCl_6$ for different initial structures, followed by 12-ns NPT simulation at 300K with the last 4 ns used for production.

| Elastic properties | Initial structure from AIMD/MLFF training[a] | Initial structure from random packing[b] |
|---|---|---|
| **Young's modulus (E)/GPa** | 2.84 ± 0.26 | 2.84 ± 1.03 |
| **Shear modulus (G)/GPa** | 1.02 ± 0.04 | 1.01 ± 0.17 |
| **Bulk modulus (B)/GPa** | 4.44 ± 0.26 | 5.05 ± 1.20 |
| **Poisson's ratio ($\nu$)** | 0.39 ± 0.04 | 0.41 ± 0.14 |

[a] From the average of the three trajectories in Table S3.

[b] A 2048-atom amorphous $LiTaCl_6$ model (256 Li, 1536 Cl, 256 Ta) was generated by geometric random packing in a 40.0 Å cubic box (64,000 Å$^3$). Ta centers were first placed by random sequential adsorption with a minimum Ta–Ta separation of 5.3 Å, yielding a low-density, maximally disordered network. Each Ta was then decorated with a rigid $TaCl_6$ octahedron (Ta–Cl = 2.37 Å) of random orientation. Interoctahedral Cl–Cl overlaps were reduced through a zero-temperature Monte Carlo optimization involving rigid-body rotations and small translations of entire octahedra while preserving octahedral geometry exactly. This lowered the number of Cl–Cl contacts shorter than 3.0 Å from 637 to 51, with all remaining contacts lying between 2.84 and 3.0 Å. Li atoms were subsequently inserted by rejection sampling into interstitial pockets of the Cl framework, subject to local constraints of Li–Cl ≥ 2.2 Å, Li–Ta ≥ 3.2 Å, Li–Li ≥ 2.8 Å, and at least four Cl neighbors within 3.2 Å. All 256 Li atoms were placed successfully. The final structure has minimum separations of Ta–Ta = 5.0 Å, interoctahedral Cl–Cl = 2.84 Å, Li–Cl = 2.2 Å, Li–Ta = 3.2 Å, and Li–Li = 2.8 Å.